\documentclass[lettersize,journal]{IEEEtran}
\usepackage{amsmath,amsfonts}
\usepackage{algorithmic}
\usepackage{algorithm}
\usepackage{array}
\usepackage[caption=false,font=normalsize,labelfont=sf,textfont=sf]{subfig}
\usepackage{textcomp}
\usepackage{stfloats}
\usepackage{url}
\usepackage{verbatim}
\usepackage{graphicx}
\usepackage{cite}
\begin{document}

\title{The Power of Fragmentation: A Hierarchical Transformer Model for Structural Segmentation in Symbolic Music Generation}

\author{Guowei Wu,
        Shipei Liu,
        Xiaoya Fan
\thanks{This paper was produced by the IEEE Publication Technology Group. They are in Piscataway, NJ.}
\thanks{Manuscript received June 26, 2022; revised June 26, 2021.}}



\maketitle

\begin{abstract}
Symbolic Music Generation relies on the contextual representation capabilities of the generative model, where the most prevalent approach is the Transformer-based model.
The learning of musical context is also related to the structural elements in music, i.e. intro, verse, and chorus, which are currently overlooked by the research community. 
In this paper, we propose a hierarchical Transformer model to learn multi-scale contexts in music. 
In the encoding phase, we first designed a Fragment Scope Localization layer to syncopate the music into chords and sections.
Then, we use a multi-scale attention mechanism to learn note-, chord-, and section-level contexts. 
In the decoding phase, we proposed a hierarchical Transformer model that uses fine-decoders to generate sections in parallel and a coarse-decoder to decode the combined music.
We also designed a Music Style Normalization layer to achieve a consistent music style between the generated sections.
Our model is evaluated on two open MIDI datasets, and experiments show that our model outperforms the best contemporary music generative models. 
More excitingly, the visual evaluation shows that our model is superior in melody reuse, resulting in more realistic music.
\end{abstract}


\section{Introduction}
Symbolic Music Generation (SMG) refers to generating continuation from the initial notes.
It has received great attention with the prosperity of deep learning~\cite{Briot-et-al:1}. 
Music can be seen as a sequence of notes in time.
A musical generative model should be able to refer to the context of note representations, as required for natural language models.
Hence, Language Models (LMs), such as the auto-regression model, are prevalently used for music generation.
A representative model is PerformanceRNN~\cite{PerformanceRNN}, an LSTM-based recurrent neural network designed to model polyphonic music with complex dynamics.
It performs well in generating short music ($\sim$ 30s).
But generating long music sequences ($\geq$ 4 minutes) is still a challenge because errors accumulate as the length of the sequence increases.
The output length is often limited in current LMs, resulting in a restricted maximum perception range.
When the length of the generated sequence exceeds the perception range, its performance degenerates quickly.

Many efforts have been devoted to maintaining long-term relevance with reasonable computational complexity in SMG tasks.
Especially the Transformer model~\cite{VaswaniSPUJGKP17} have helped the advance state-of-the-art (SOTA) in symbolic music generation.
For instance, Google researchers ~\cite{MusicTransformer} have proposed MusicTransformer, which can deal with longer sequences ($\sim$ 4096 notes) with optimized intermediate memory occupation.
MuseNet~\cite{Musenet} is a Transformer-based model, using the same general-purpose unsupervised technology as GPT-2~\cite{GPT-2}, that generates 4-minute music and discovers patterns of harmony, rhythm, and style in long-form music.
Despite recent improvements in long-term music generation, existing approaches fail to learn music structure effectively. 
This may be attributed to their incompetence in segmenting structural elements.

Most music is typically structural, such as intro, verse, chorus, etc., meaning that some attractive melodies are repeated throughout the song.
In traditional neural networks, hierarchical architectures have been proposed to generate these structured melodies.
MusicVAE~\cite{LVM} uses a hierarchical decoder to model structural elements, such as the repetition and variation between measures and sections of a piece of music.
MuseGAN~\cite{MuseGAN} consists of three Generative Adversarial Networks (GANs) models, called jamming model, composer model, and hybrid model, for generating multi-track music with their respective temporal dynamics.
Researchers have shown that the hierarchical architecture enables the extraction of structural context, enhancing the long-term relevance in the decoding phase.
A similar idea has also been introduced to a transformer-based model for text generation task ~\cite{GabrielBSBFC21} that references paragraph-level context and achieves SOTA performance.
Therefore, we argue that structural segmentation and section-level contextual learning are the primary challenges in generating realistic long-term music.

\begin{figure*}[htbp]
\centering
\includegraphics[width=1.96\columnwidth]{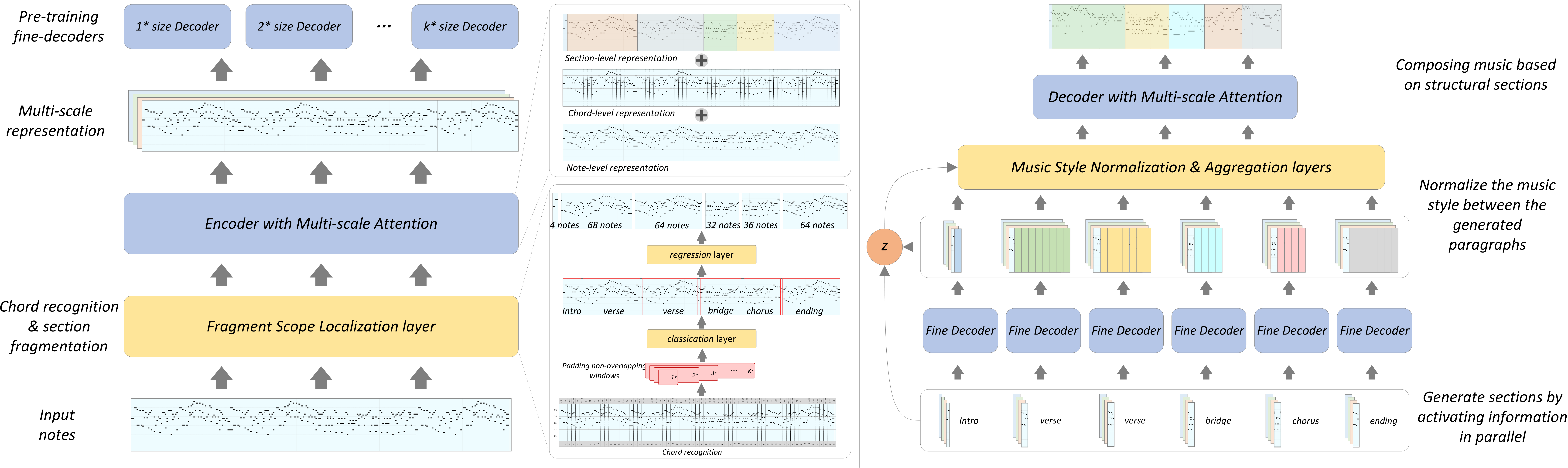}
\centerline{\footnotesize(a) Encoding \quad\quad\quad\quad\quad\quad\quad\quad\quad\quad\quad\quad\quad\quad\quad\quad\quad\quad\quad\quad \footnotesize(b) Decoding}
\caption{An overview of the proposed framework. 
(a) The encoding phase: 
The FSL layer is used to recognize chords and sections.
It consists of two sub-layers, a classification (cls) layer and a regression (reg) layer, for recognizing sections and locating the location of each recognized section, respectively.
Then, these structural elements are fed into a Transformer encoder that uses multi-scale attention to embed musical representations at the note, chord and section levels.
Fine-decoders of different lengths are subsequently pre-trained by multi-scale representations.
(b) The decoding phase: 
At the bottom of the hierarchical architecture, fine-decoders are used to generate output sections in parallel.
We introduced an MSN layer to normalize the music style through a variable $z$ and maintain consistency between the generated sections.
The normalized sections are combined by an aggregation layer and fine-tuned by a followed coarse-decoder.
}
\label{fig1}
\end{figure*}

In this paper, we propose a hierarchical Transformer model to learn multi-scale representations extracted by structural segmentation methods.
Concretely, chords were first recognized from note sequence and then fed to a custom-designed Fragment Scope Localization (FSL) layer.
The note sequence was segmented into several structural sections by the region proposal method with two sub-layers, namely the cls layer and the reg layer.
Then, fine-decoders were pre-trained with multi-scale attention through these structural corpora, enabling chord-level and section-level contextual learning. 
Next, the first few notes and labels are input as activation information to generate sections.
We propose a Music Style Normalization (MSN) layer to control the music style of the generated sections.
The followed aggregation layer and coarse-decoder are used to combine the generated sections and fine-tune them at the global scale, respectively.
An illustrative diagram of the proposed framework is shown in Fig. \ref{fig1}. 
Our main contributions are as follows:

\begin{itemize}
\item We designed an FSL layer that divides the music into structural elements and further analyzed its utility with different settings through ablation studies.
\item We proposed a multi-scale attention mechanism to learn music representations at the note, chord, and section levels.
Visual evaluation shows its superiority in reusing melodies, such as chord progression at the chord-level.
\item We proposed a hierarchical architecture, using fine-decoders to generate sections in parallel and fine-tune the combined music with a coarse-decoder.
Since the proposed model generates the music by section, it can alleviate the problem of error accumulation. 
Experimental results show that it achieves SOTA performance on two open datasets.
\item We designed an MSN layer to control music style and argued for the importance of Mutual Information (MI) in this process.
Thus, the proposed model showed superiority in the style consistency of generated music compared to other models.
\end{itemize}

\section{Related works}
\subsection{Transformer baselines}
The Transformer model has been the typical choice for sequence data modeling due to its advantages in contextual representation capabilities.
However, applying the Transformer model to the SMG task is computationally prohibitive since the computational complexity increases quadratically with the sequence length ~\cite{li2019enhancing}.
Reformer ~\cite{Reformer} is an effective work, where the dot-product attention is replaced by a locality-sensitive hashing-based calculation, reducing the complexity from $O\left(L^{2}\right)$ to $O(L \sqrt{L})$.
Many other Transformer models have been devoted to resolving the problem of building longer-term dependencies in sequence prediction tasks.
For instance, Transformer-XL~\cite{Transformer-XL} learns dependencies beyond a fixed length by a segment-level recurrence mechanism and a positional encoding scheme.
Some approaches introduce additional modules to improve music generation.
Transformer-GANs~\cite{Transformer-GANs} introduces a pre-trained discriminator that uses adversarial loss to complement the negative log-likelihood objective, enabling improvement in synthesizing minute-long compositions.
These proposed baselines provide the feasibility of solving SMG tasks by Transformer models.

\subsection{Music structure segmentation}
Structure segmentation is commonly used for modeling long-form music, which can be tackled with hierarchical architecture.
In a music annotation work ~\cite{WangSW19}, attentive convolution networks and recurrent networks were hierarchically combined to solve the problem of audio music representation and structure learning.
McCallum ~\cite{Segmentation} explores the use of Convolutional Neural Networks (CNNs) for unsupervised training in music segmentation, aiming to detect the boundaries of musical fragments in audio music.
Dai ~\cite{Hierarchical:2} proposed a hierarchical model to generate a full-length melody guided by long-term repetitive structure and achieved near-human performance in melody generation about half the time.
These works suggest that it is beneficial to design a structural segmentation module for SMG tasks.

\subsection{Music Style Control}
Instance normalization methods are often used for style transfer.
For instance, Huang et al. ~\cite{HuangB17} proposed an adaptive instance normalization approach to achieve flexible style control. 
Ling et al. ~\cite{RA_AIN} used a region-aware adaptive instance normalization module to formulate the visual style of the background and transfer it to the foreground.
MI is also commonly used for style control.
Chawla ~\cite{ChawlaY20} proposed a model for formality style transfer, which maximized the MI between original and target styles as the training objective and achieved better performance.
Inspired by these studies, we utilized a MI-based strategy to maintain a consistent style in long-form music.

\section{The proposed model}
The key challenge for the SMG task is to generate long-term music with structural relevance and consistent style.
We begin with a description of note sequence encoding and structural element fragmentation, followed by an introduction of the hierarchical architecture and music style normalization. 
A theoretical analysis of the coverage of attention patterns of the proposed and baseline models was also performed.

\subsection{Encoding: Fragmentation scope localization and Multi-scale attention mechanism}
During the encoding phase, the FSL layer is used to recognize chords and sections, which facilitates the extraction of multi-scale contexts in music.
Then, we propose multi-scale attention to learn musical representations at different time scales and pre-train the fine-decoders.
The detail of the FSL layer and multi-scale attention is illustrated in Fig. \ref{fig1} (a).

\textbf{Chord recognition:} 
A chord, in music, is a group (typically three or more) of notes that sound together and serve as a basis of harmony.
In tonal music rules, a chord can also be a group of notes that appear dispersed in a bar.
Therefore, we extracted the chord feature, called chord profile, by counting the notes of each bar.
The notes (transferred to the same octave) was mapped to a 12-dimensional vector, usually called twelve-tone equal temperament, which is a set of pitch classes $\left[\textbf{C}_{1}, Db_{2}, \textbf{D}_{3}, Eb_{4}, \textbf{E}_{5}, \textbf{F}_{6}, F\sharp_{7}, \textbf{G}_{8}, Ab_{9}, \textbf{A}_{10}, Bb_{11}, \textbf{B}_{12}\right]$.

\renewcommand\arraystretch{1.2}
\begin{table}[htbp]
\centering
\caption{Description of usual chord profiles.}
\label{table1}
\resizebox{0.98\columnwidth}{!}{
\begin{tabular}{lcc|lcc}
\hline
\multicolumn{1}{c}{\textbf{Id}}&\textbf{Chord type}&\textbf{Chord profile}&\textbf{Id}&\textbf{Chord type}&\textbf{Chord profile}\\
\hline
1&$\left(empty\right)$&[0, 4, 7]&25&79$\sharp$&[0, 3, 4, 7, 10]\\
2&+&[0, 4, 8]&26&79$\sharp$11$\sharp$&[0, 3, 4, 6, 7, 10]\\
3&+7&[0, 4, 8, 10]&27&79$\sharp$13&[0, 3, 4, 5, 7, 9, 10]\\
4&+79&[0, 2, 4, 8, 10]&28&7911&[0, 2, 4, 5, 7, 10]\\
5&+79$\sharp$&[0, 3, 4, 8, 10]&29&7911$\sharp$&[0, 2, 4, 6, 7, 10]\\
6&+7911$\sharp$&[0, 2, 4, 6, 8, 10]&30&7913&[0, 2, 4, 5, 7, 9, 10]\\
7&+79b&[0, 1, 4, 8, 10]&31&7913b&[0, 2, 4, 5, 7, 8, 10]\\
8&+j7&[0, 4, 8, 11]&32&79b&[0, 1, 4, 7, 10]\\
9&-&[0, 3, 7]&33&79b13&[0, 1, 4, 5, 7, 9, 10]\\
10&-6&[0, 3, 7, 9]&34&79b13b&[0, 1, 4, 5, 7, 8, 10]\\
11&-69&[0, 2, 3, 7, 9]&35&7alt&[0, 1, 4, 6, 10]\\
12&-7&[0, 3, 7, 10]&36&j7&[0, 4, 7, 11]\\
13&-79&[0, 2, 3, 7, 10]&37&j79&[0, 2, 4, 7, 11]\\
14&-7911&[0, 2, 3, 5, 7, 10]&38&j79$\sharp$&[0, 3, 4, 7, 11]\\
15&-7913&[0, 2, 3, 5, 7, 9, 10]&39&j79$\sharp$11$\sharp$&[0, 3, 4, 6, 7, 11]\\
16&-79b&[0, 1, 3, 7, 10]&40&j7911$\sharp$&[0, 2, 4, 6, 7, 11]\\
17&-j7&[0, 3, 7, 11]&41&m7b5&[0, 3, 6, 10]\\
18&-j7911$\sharp$&[0, 2, 3, 6, 7, 11]&42&o&[0, 3, 6]
\\
19&-j7913&[0, 2, 3, 5, 7, 9, 11]&43&o7&[0, 3, 6, 9]\\
20&6&[0, 4, 7, 9]&44&sus&[0, 5, 7]\\
21&69&[0, 2, 4, 7, 9]&45&sus7&[0, 5, 7, 10]\\
22&6911$\sharp$& [0, 2, 4, 6, 7, 9]&46&sus79&[0, 2, 5, 7, 10]\\
23&7&[0, 4, 7, 10]&47&sus7913&[0, 2, 5, 7, 9, 10]\\
24&79&[0, 2, 4, 7, 10]&48&Non-chord&[0]\\
\hline
\end{tabular}}
\end{table}

We counted the 47 most common forms of chord composition (major triads, minor sixth chords, minor-major seventh chords, suspension chords, etc.).
Single notes that cannot form a chord are denoted as type 48.
For example, the $C$ triad chord consists of three notes: a root note $C$, intervals of a third $E$, and a fifth above the root note $G$.
Thus, the $C$ triad chord can be represented by $\left[C, E, G\right]$, and in twelve-tone equal temperament by $C_{chord}=C_{1}+Triad_{\left[0, 4, 7\right]}=\left[C_{1}, E_{5}, G_{8}\right]$.
The chord profile used in our model is shown in Tab. \ref{table1}. 

\begin{figure}[htbp]
\centering
\includegraphics[width=0.96\columnwidth]{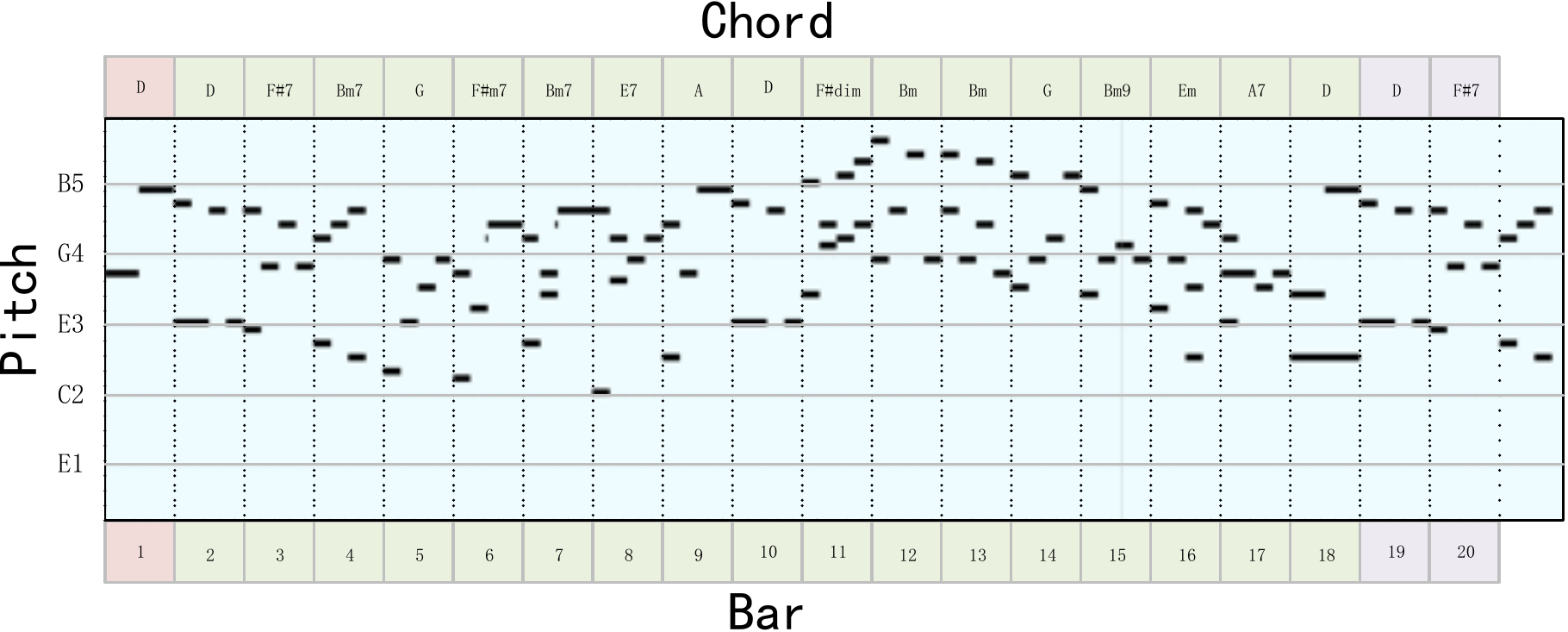}
\caption{
Example of chord recognition results.
The first 20 music bars are displayed in a visual MIDI diagram, which can easily exhibit the note constitution of the chord between bar lines.
}
\label{fig2}
\end{figure}

In our model, chords are identified automatically from MIDI data in two steps: 
1. Template matching for complete chord; 
2. Sequence estimation for unidentified bars.
More specifically, we use the template matching method to match the same chords as in the chord template, including the 47 common chord profiles.
These matching chords have neither missing notes nor other ornamental notes.
Then, we construct an n-gram sequence to represent unidentified bars and use a nonlinear classifier to calculate correlation, which can determine the probability of the chord category.
The result of chord recognition is demonstrated in Fig. \ref{fig2}.

A series of chords is called a chord progression, and it has the characteristic of being periodically reused in sections of the same label.
For example, in Fig. \ref{fig2}, the 19, 20, and 21 bars use the same chord as the 2, 3, and 4 bars.
The construction of these specific chord progressions is one of the goals we hope to achieve with our model.

\textbf{Fragment region proposal:} 
We use the region proposal method to selectively search for music sections.
The candidate windows $\omega$ used to extract section location and length, is applied to the note sequence $chord\left(x\right)$ labeled by the chord profiles.
Thus, the window size is set to a multiple of the chord length (usually 16 semiquaver notes), with $K$ sizes available.
Normally, sections are extracted through windows sliding from left to right. 
The left-to-right (L2R) strategy can be written as:
\begin{equation}
Cand\left[m\right]=\omega\left[k\right] \cdot chord\left(Loc_{m}: Loc_{m}+\omega\left[k\right]\right)
\label{eq1}
\end{equation}
where $Cand\left[m\right]$ denotes the scope of $m$-th extracted fragment;
$\omega\left[k\right]$ denotes the candidate window with the appropriate size $k$;
The current location is calculated by $Loc_{m}=\sum_{j=1}^{m-1}\omega_{j}$;
The goal of L2R strategy is to select suitable windows with the maximum classification probability $\mathop{\arg\max}\limits \left(cls\left(\omega\cdot chord\left(x\right)\right)\right)$ in order from left to right. 

\begin{figure}[htbp]
\centering
\includegraphics[width=1\columnwidth]{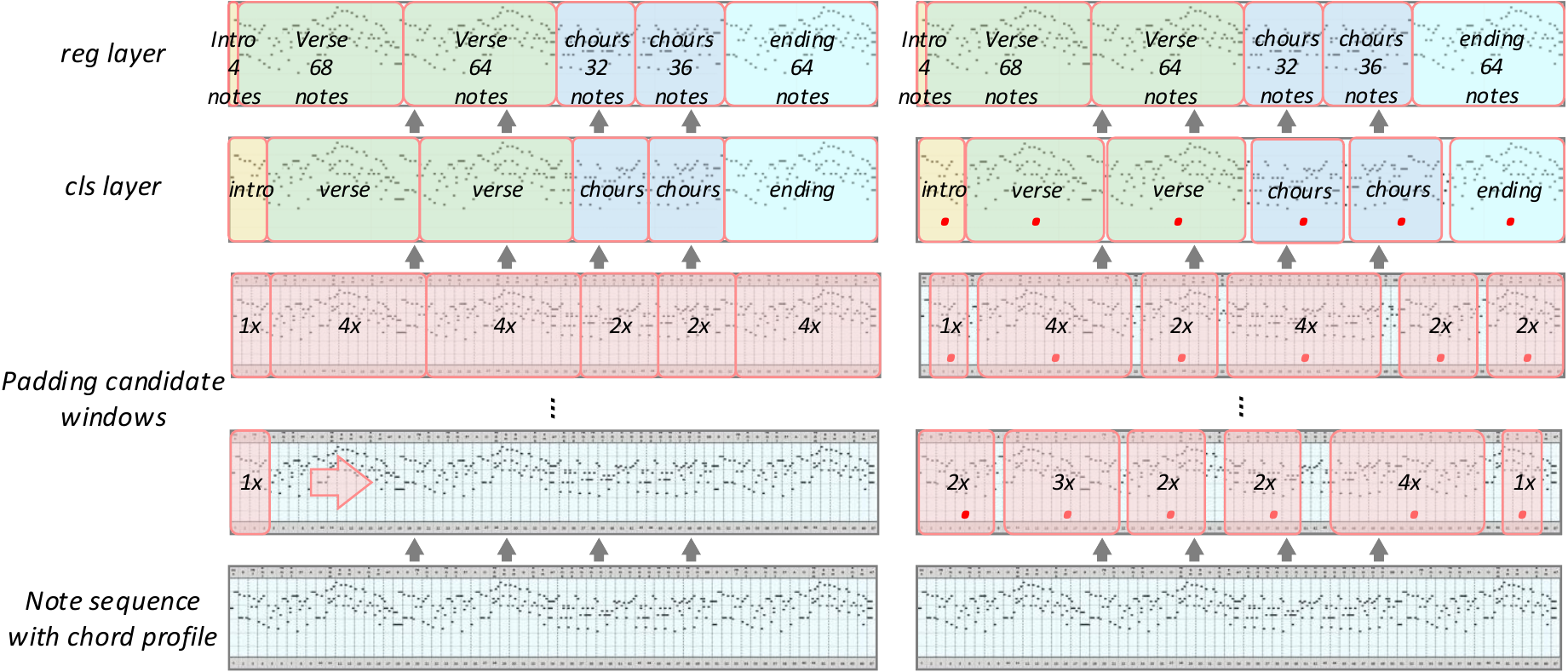}
\centerline{\footnotesize(a) Left-to-right strategy \quad\quad\quad \footnotesize(b) Global strategy}
\caption{Two alignment strategies for candidate window.
(a) The candidate windows slide over the note embeddings to produce proposal regions.
The cls layer identifies the section categories, and the reg layer locates the section scopes to the ground truth.
(b) The central locations of the candidate scopes are randomly initialized on the note embedding.
Then, the cls layer identifies the section labels, and the reg layer regresses the candidate scopes.
}
\label{fig3}
\end{figure}

Obviously, the L2R strategy is prone to error accumulation of the window location. 
To deal with this, we design a global strategy, where a random location $\Delta Loc$ is added to the input, and the non-overlapping windows are padded.
The global strategy can be written as:
\begin{equation}
Cand^{*}\left[m\right]= \omega\left[k\right] \cdot chord\left(\Delta Loc_{m}\pm\frac{1}{2}\omega\left[k\right]\right)
\label{eq2}
\end{equation}
where $\Delta Loc_{m}$ denotes the center of the $m$-th random location;
The non-overlapping windows are padded by maximizing the average confidence, calculated by $\mathop{\arg\max}\limits \left(\frac{1}{M}\sum_{m=1}^{M} cls \left(\omega \cdot chord\left(x\right) \right)\right)$.
The comparison of the two alignment strategies is shown in Fig. \ref{fig3}.

\textbf{Candidate classification:}
The cls layer produces category probabilities $p_{\varphi} \in \mathbb{R}^{\phi}$ to predict the label $\varphi \in \phi$ of candidate sections, such as intro, verse, chorus.
The number of labels in the training data is unbalance, so we use Gumbel Softmax~\cite{KusnerH16} to optimize the cls layer, which is defined as:
\begin{equation}
L_{cls}=-\log \frac{\exp \left(\left(p_{\varphi} + \theta_{\varphi}\right)/ \tau\right)}{\sum_{\tilde{\varphi} \sim \phi} \exp \left(\left(p_{\tilde{\varphi}}+ \theta_{\tilde{\varphi}}\right)/ \tau\right)} \label{eq3}
\end{equation}
where $\tau$ is a non-negative coefficient of spread in the $Gumbel$ distribution;
The smaller the coefficient, the closer the sample expectation is to the $Argmax$ function;
The larger the coefficient, the more average the sample expectation is. 
$\theta$ are samples drawn from $Gumbel\left(0,1\right)$.
The $\tilde{\varphi}$ lists the all labels, and $p_{\tilde{\varphi}}$ is the probabilities of candidate section in every label.
The value of $L_{cls}$ should be close to 0 if the candidate is consistent with the ground-truth label.

\textbf{Scope regression:} 
To measure the overlap ratio between the candidate and ground-truth sections, we check the percentage of shared and different notes in these paired sections.
Let $D=\left[(\omega_{1},y_{1}),\cdots,(\omega_{M},y_{M})\right]$ be a training set of instance pairs, $y$ are the ground-truth sections. 
We use the Jaccard similarity coefficient to evaluate the similarity between the predicted and ground-truth scopes, which is denoted as $Jac\left(a, b\right)=\left(a \cap b\right)/\left(a \cup b\right)$.
The regression loss function can be calculated by:
\begin{equation}
L_{reg}=-\log (Jac\left(\omega_{m},y_{m}\right))
\label{eq4}
\end{equation}
A low regression loss indicates high coincidence between the prediction and the ground-truth. 
Accordingly, the loss function of our FSL layer can be represented by:
\begin{equation}
L_{FSL}=\frac{1}{N_{cls}} \sum_{m=1}^{N_{cls}} L_{cls}+\frac{1}{N_{reg}} \sum_{m=1}^{N_{reg}} L_{reg}\label{eq5}
\end{equation}
where $L_{cls}$ and $L_{reg}$ are normalized by $N_{cls}$ and $N_{reg}$, which are the sample number in these two layers, respectively.

After passing through the FLS layer, the note sequence is segmented into structural sections.
The fragmentation process is formulated as:
\begin{equation}
Section_{m}\left(x\right)= reg\left(cls\left(Cand\left[m\right]\right) \right)\label{eq6}
\end{equation}
where $cls$ and $reg$ stands for the classification and regression operation of the $m$-th fragment, respectively;

\begin{figure}[htbp]
\centering
\includegraphics[width=0.98\columnwidth]{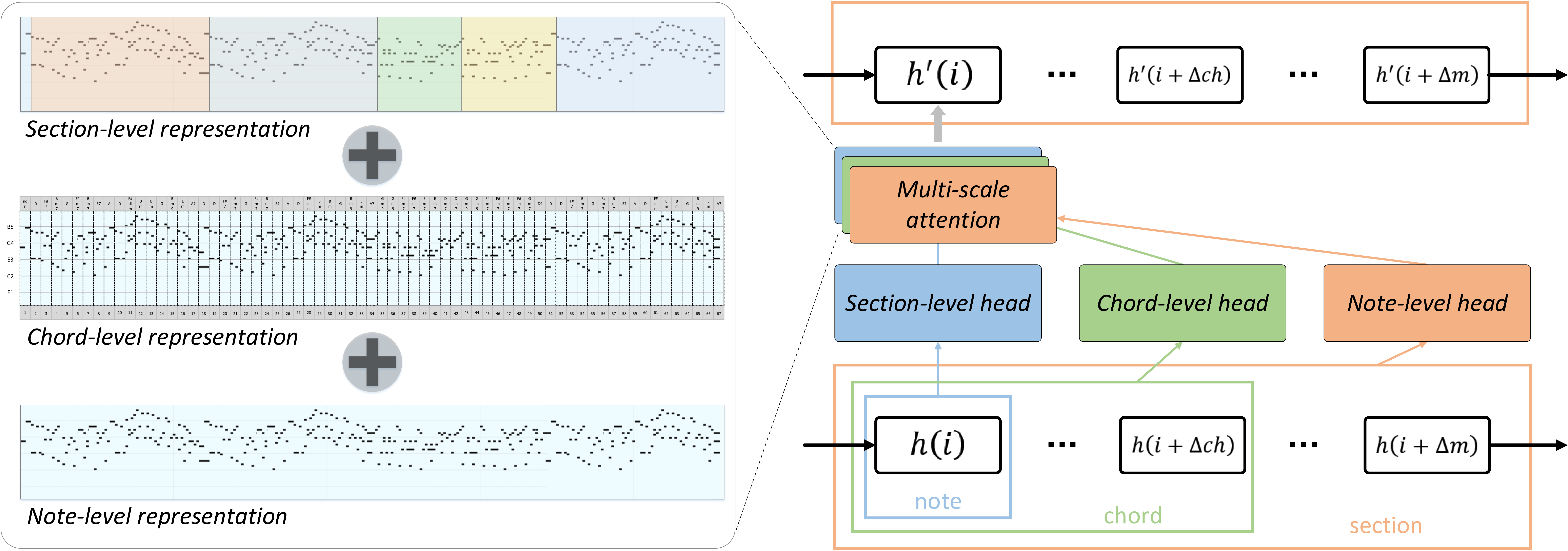}
\caption{
A diagram of multi-scale attention.
Each layer in the encoder consists of three groups of attention heads, capable of learning music representations at different scale.
}
\label{fig4}
\end{figure}

\textbf{Multi-scale attention and fine-decoders pre-training:} 
In Transformer models, the attention pattern determines the perceptual range.
We designed a multi-scale attention mechanism for musical representation, to enable learning of chord-level and section-level contextual features, as shown in Fig. \ref{fig4}.

Given a vector $X\in \mathbb{R}^{L\times D}$ of note sequence, where $L$ is the length of section sequence and the $D$ is the dimension of the vector.
The output embedding $h\left(x\right)$ is obtained by a Transformer encoder, which has $l$ hidden layers, denoted as:
\begin{equation}
h_{l}=LN\left(h_{l-1}+FFN\left(MS\_Attention\left(Q,K,V\right)\right)\right)
\label{eq7}
\end{equation}
where $Q$, $K$, $V$ represent the Query, Key, Value vector in the Transformer-based model;
$LN$ and $FFN$ represent the layer normalization operation and feed-forward network, respectively;
$MS\_Attention$ represents the multi-scale attention.

Multi-scale attention has a scale pattern $S=\left(S_{1}, \cdots, S_{i}\right)$, which we parameterize by default three scales $S_{note}$, $S_{chord}$, and $S_{section}$, to control its working scope.
The multi-scale attention can be calculated as:
\begin{equation}
\begin{aligned}
& MS\_Attention=\left[\begin{array}{c}
head_{1}\left(S_{1}\right),\cdots,head_{N_{1}}\left(S_{1}\right)\\
\cdots\\
head_{1}\left(S_{i}\right),\cdots,head_{N_{i}}\left(S_{i}\right)\end{array}\right]
\end{aligned}
\label{eq8}
\end{equation}
where $N_{i}$ is the number of heads at each scale;
The attention head is the weighted sum of scaled dot-product of the input vectors, calculated by:
\begin{equation}
head\left(S_{i}\right)=Softmax\left(\frac{Q_{S_{i}} K_{S_{i}}^{T}}{\sqrt{D}}\right) \cdot V_{S_{i}}
\label{eq9}
\end{equation}
\begin{equation}
\begin{aligned}
& \left[\begin{array}{c}
Q_{S_{i}}\\
K_{S_{i}}\\
V_{S_{i}}\end{array}\right]=
\left[\begin{array}{c}
W_{Q}\\
W_{K}\\
W_{V}\end{array}\right]\odot 
X\left(S_{i}\right)
+\left[\begin{array}{c}
0\\
P_{S_{i}}\\
P_{S_{i}}\end{array}\right]
\end{aligned}
\label{eq10}
\end{equation}
where $W_{Q}, W_{K}, W_{V}$ represents the weight matrix that transform the input into the $Query$, $Key$, and $Value$ vector, respectively; 
$X(S_{i})$ defines the references at the $i$ scale, corresponding to $note(x)$, $chord(x)$ and $section(x)$.
$P_{S_{i}}$ represent the position embeddings (PEs) in each scale, for example, use relative PEs (RPEs) in the note and chord scales, and absolute PE (APEs) in the section scale.

Through the $MS\_Attention$, we pre-trained several fine decoders, whose maximum length is according to the section length.
The training steps and important parameters are detailed in Appendix.

\subsection{Decoding: hierarchical architecture with Music Style Normalization}
The proposed hierarchical model, a bottom-up architecture, aims to generate notes at scales from fine to coarse.
The decoding phase is shown in Fig. \ref{fig1} (b).

At the bottom of the hierarchical architecture, we use the fine decoders, pre-trained with different maximum lengths, to generate music sections.
Each section is decoded in parallel with an activation information, which usually uses the first few notes and a label:
\begin{equation}
G_{m}=Decoder_{fine}\left(Section_{m}\left(x\right)\left[0:r\right] ; z_{m}\right)
\label{eq11}
\end{equation}
where $G_{m}$ represents the $m$-th generated section;
Fine-decoders are activated by start notes $Section_{m}\left(x\right)$ with default length $r$;
$z_{m}$ indicates the target label of the generated section, i.e., intro, verse, and chorus, etc.

Specifically, given the start notes $\left(h^{\prime}_{1}, \cdots,h^{\prime}_{r}\right)$, the output embedding $h^{\prime}_{r+1}$ is obtained by:
\begin{equation}
h^{\prime}_{r+1}=LN\left(h^{\prime}_{r}+FFN\left(a_{r+1} \left( h^{\prime}_{1}:h^{\prime}_{r} \right)\right)\right)
\label{eq12}
\end{equation}
where $a_{n}$ is the self-attention block, which calculates a weight score from previous notes.

\textbf{Music Style Normalization:} 
To enhance the style consistency between the original and generated section, we designed an MSN layer.
The musical style is quantified by a variable $z$ that has the same dimensions as the note embedding.
Through the MSN layer, the style variable $z$, initialized by the section label, is transferred to the generated sections, which can be expressed as $z_{m} \to z_{m}^{\prime}$.
For example, the MSN layer normalizes the mean and standard deviation of note pitches of each original section and maps them to the generated section:
\begin{equation}
MSN\left(G,z\right)=\gamma_{z}\left(\frac{G_{m}-\mu(z_{m}^{\prime})}{\sigma(z_{m}^{\prime})}\right)+\beta_{z}
\label{eq13}
\end{equation}
where $\gamma_{z}$, $\beta_{z}$ are the scaling and translation parameters, respectively, which are calculated independently for each category.
$\mu(z_{m}^{\prime})$ and $\sigma(z_{m}^{\prime})$ represent the mean and standard deviation of note pitches in $i$-th generated section, respectively.

At the up of the hierarchical architecture, these sections are aggregated into a long sequence by the weighted concatenation operation $v$.
The coarse-decoder used to learn the global contextual reference from the merged sequence.
The coarse decoding process can be represented by:
\begin{equation}
G=Decoder_{coarse}\left(v\left(MSN\left(\sum_{m=1}^{M}G_{m}\right)\right)\right)\label{eq14}
\end{equation}
where $G$ stands for the output sequence;
The coarse-decoder $Decode_{coarse}$ uses a multi-scale attention with same fixed-length as the encoder.

\textbf{Multi-Task Loss:} 
The loss function we used in decoding phase consists of two terms: 
(1) The $L_{MLM}$ for the note prediction in coarse decoder;
(2) The $L_{style}$ for music style normalization.
\begin{equation}
L_{Decoding}=L_{MLM}\left(h_{\Pi} \mid h_{-\Pi}\right)+\lambda L_{style}\left(x, G\left(x,z\right)\right)
\label{eq15}
\end{equation}
where hyperparameter $\lambda$ is used to balance the magnitude of loss terms. 

Given a sequence $h_{1:n}$, decoders predicts $T$ masked notes among them.
We minimize the following MLM loss:
\begin{equation}
L_{\mathrm{MLM}}\left(h_{\Pi} \mid h_{-\Pi}\right)=-\frac{1}{T} \sum_{t=1}^{T} \log p\left(h_{\Pi_{t}} \mid h_{-\Pi}\right)
\label{eq16}
\end{equation}
where $h_{\Pi}$ and $h_{-\Pi}$ denotes the masked and unmasked notes, respectively.

To improve the stylistic similarity between the original and generated sections, we maximize their MI in the optimizer.
Mathematically, we can examine the $KL$ divergence between the joint probability distribution and the marginal probability distribution to determine whether the two variables are approximately independent. 
Thus, we can use the variational information maximization method ~\cite{MIboundary} to estimate the $I\left(x;G\left(x,z\right)\right)$ by instantiating an intermediate tensor $Q\left(x \mid G\left(x,z\right)\right)$:
\begin{equation}
\begin{array}{c}
I(x ; G(x, z))=E_{p(x, G(x, z))}\left[\log \frac{Q(x \mid G(x, z))}{P(x)}\right]\\
+E_{p(G(x, z))}\left[D_{K L}(P(x \mid G(x, z)) \| Q(x \mid G(x, z)))\right]
\end{array}
\label{eq17}
\end{equation}
where $P\left(x\right)$ is the distribution of training data;
$P\left(G\left(x,z\right)\right)$ is the distribution of generated samples;
We empirically deduce that $D_{K L}>0$, then the $I\left(x;G\left(x, z\right)\right)$ can transform to:
\begin{equation}
I(x ; G(x, z)) \geq E_{p(x, G(x, z))}\left[\log Q(x \mid G(x, z))\right]+H\left(X\right)
\label{eq18}
\end{equation}
where $H\left(x\right)$ is the differential entropy of $X$. 
We can derive an initial lower bound, which is tight when $Q\left(x \mid G\left(x,z\right)\right)=P\left(x \mid G\left(x,z\right)\right)$.
The musical style loss $L_{style}\left(x,G\right)$ can calculate on $I\left(x;G\left(x,z\right)\right)$:
\begin{equation}
\begin{array}{c}
L_{style}(x, G)=E_{p(x, G(x, z))}[\log Q\left(x \mid G\left(x,z\right)\right)]+H(z) \\
=E_{x \sim P(x)}\left[E_{G^{\prime}\left(x,z\right) \sim P(G\left(x,z\right) \mid x)}\left[\log Q\left(x \mid G^{\prime}\left(x,z\right)\right)\right]\right] \\+H(z)
\leq I(x ; G(x, z))
\end{array}
\label{eq19}
\end{equation}

The $L_{style} \left(x,G\right)$ can be thought of as a negative reconstruction error, in which the gradient of intermediate tensor $Q\left(x \mid G\right)$ is tractable. 
Therefore, $L_{style}$ can be maximized via the re-parameterization trick. 
This optimization conforms to the information-theoretic regularization: MI is high when the two musical sections are similar.
Hence, it is feasible to add the $L_{style}$ loss to maintain consistency of generated sections.

\textbf{Attention pattern analysis:} 
At last, we analyze the advantage of the proposed hierarchical model in attention patterns and compare it with others.
We use a 16*16 image to compare attention mechanisms, where the differences between the four patterns are revealed by the connection matrix of the output (rows) and the input (columns).

\begin{figure}[htbp]
\centering
\includegraphics[width=0.98\columnwidth]{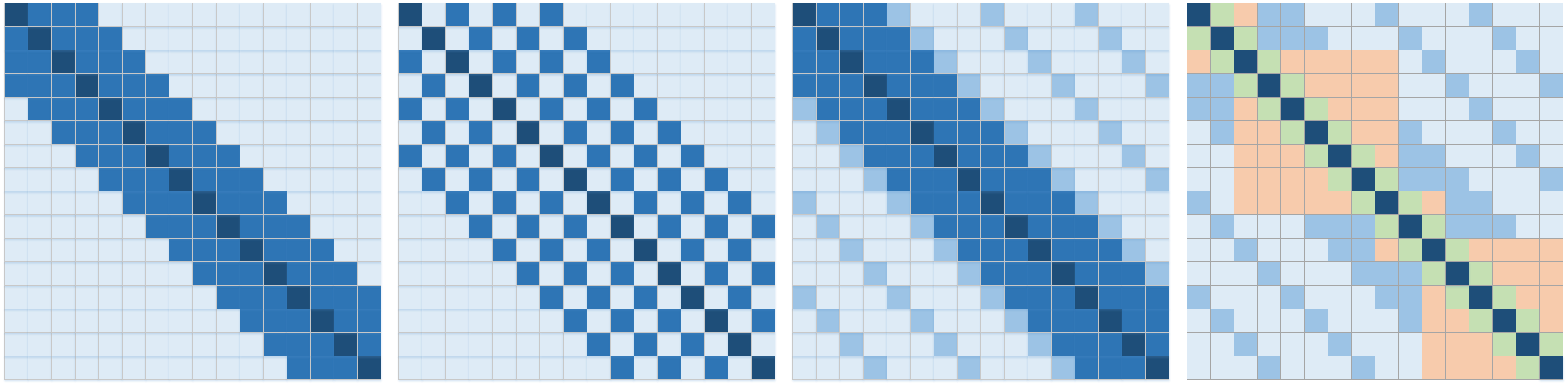}
\centerline{\footnotesize(a) local \quad\quad\quad\ \footnotesize(b) dilated \quad\quad\quad \footnotesize(c) sparse \quad\quad\quad \footnotesize(d) hierarchical}
\caption{Illustration of four attention patterns: 
(a) Local attention (window size 3); 
(b) Dilated attention (dispersion stride 1); 
(c) Sparse attention (window size 3 and dispersion stride 3); 
(d) multi-scale attention (4 fine-decoders for 4 sections of length $[3, 7, 3, 6]$);
The depth of blue represents the weight of attention, while the white color block indicates the absence of attention.
The light blue, green, and orange block correspond to the attention range of note-, chord- and section-level, respectively.
}
\label{fig5}
\end{figure}

As shown in Fig. \ref{fig5} (a), local attention cannot exchange information with distant notes.
This situation is the opposite of dilated attention, see Fig. \ref{fig5} (b).
The drawback of sparse attention, in Fig. \ref{fig5} (c), is that the dilated stride is fixed and has no structure-related connection.
In Fig. \ref{fig5} (d), our multi-scale attention enables both intra- and inter-section connectivity, which has the advantage of enhancing intra-section integrity while learning section-level dependencies.
The range and location of section-level attention are adaptive to the corresponding section in each music composition rather than hard-coded in advance. 
Therefore, section segmentation can be viewed as a dynamic assignment of full attention at a local scale, which has decreased computational cost while maintaining the long perception range of full attention.

\renewcommand\arraystretch{1.2}
\begin{table*}[!htbp]
\centering
\caption{Comparison of model performance with four other models, tested on J.S. Bach and Maestro datasets.
For each metric, the closer to the training set the values is, the better.
The values of the metrics are expected to be close to that for the training set.
The best performance is highlighted in bold.
}
\label{table2}
\resizebox{2\columnwidth}{!}{
\begin{tabular}{|c|c|c|c|c|c|c|c|c|c|c|c|c|}
\hline
\textbf{Models}&\textbf{PPL}&\textbf{PCU}&\textbf{ISR}&\textbf{PRS}&\textbf{TUP}&\textbf{PR}&\textbf{APS}&\textbf{IOI}&\textbf{PCH}&\textbf{GS}&\textbf{CPI}&\textbf{SI}\\
\hline
Training Set (J.S. Bach)&-&8.407&0.835&0.422&25.45&34.32&11.67&0.064&2.994&0.930&0.985&0.011\\
\hline
PerformanceRNN&1.96&\textbf{7.5}&0.853&0.446&9.3&28.4&3.901&0.05&2.665&0.999&0.999&0\\
Transformer-XL&1.846&5.742&0.785&0.383&34.859&40.136&10.763&0.074&2.531&0.919&0.990&0.008\\
MusicTransformer&1.833&5.768&0.792&0.386&36.069&42.542&10.257&0.075&2.535&0.916&0.993&0.009\\
Transformer-GANs&1.789&5.9&0.807&\textbf{0.423}&34.817&40.41&\textbf{10.993}&\textbf{0.072}&2.537&\textbf{0.921}&\textbf{0.987}&0.009\\
Hierarchical model (Ours)&\textbf{1.716}&7.061&\textbf{0.819}&0.412&\textbf{25.194}&\textbf{34.597}&10.982&0.124&\textbf{2.944}&0.872&0.994&\textbf{0.01}\\ 
\hline
\hline
Training Set (Maestro v3.0.0)&-&6.665&0.815&0.429&66.01&68.15&11.59&0.083& 2.886&0.901&0.958&0.208\\
\hline
PerformanceRNN&1.403&3.99&0.682&0.268&5.52&10.71&3.98& 0.099&1.912&0.880&0.322&0.014\\
Transformer-XL&1.188&6.121&0.796&0.305&53.53&61.54&11.18& 0.067&2.639&0.931&\textbf{0.957}&0.083\\
MusicTransformer&1.167&6.16&0.833&0.499&55.55&62.72&\textbf{11.69}&0.071&2.539&0.919&0.990&0.095\\
Transformer-GANs&1.145&6.136&0.814&0.298&56.17&63.93&11.95&0.091&3.51&0.919&0.997 &0.129\\
Hierarchical model (Ours)&\textbf{1.063}&\textbf{6.716}&\textbf{0.815}&\textbf{0.429}&\textbf{57.26}&\textbf{67.62}&11.35&\textbf{0.087}&\textbf{2.985}&\textbf{0.894}&0.987&\textbf{0.191}\\
\hline
\end{tabular}}
\end{table*}

\section{Experiment Results}
We used the same methods as MusicTransformer to process MIDI data.
The structural information of musical sections is automatically extracted from the music score, which are obtained by the GuitarPro software, a music recording tool. 
The annotation extraction method is detailed in Appendix.

\textbf{Datasets:} 
For pre-training the FSL layer, we collected a total of 300 scores with 3,188 sections, called the GuitarPro dataset.
After pre-training the FSL layer, we trained and tested the proposed hierarchical Transformer based on two canonical music corpora:
(1) J.S. Bach Chorales ~\cite{DeepBach} for short-term generation;
(2) Maestro v3.0.0 ~\cite{MAESTRO} for long-term generation.
The sample length of the Maestro dataset is significantly longer than that of the J.S. Bach dataset.
The split standard of these datasets abides by the rule of 80/10/10 proportion for train/validation/test. 
The detailed information of datasets is shown in Tab. \ref{table3}.

\renewcommand\arraystretch{1.2}
\begin{table}[htbp]
\centering
\caption{Description of datasets.
}
\label{table3}
\resizebox{0.98\columnwidth}{!}{
\begin{tabular}{|c|c|c|c|c|}
\hline
\textbf{Dataset}&\textbf{Scores}&\textbf{Sections}&\textbf{Notes}&\textbf{Average length}\\
\hline
GuitarPro (Ours)& 300 &3,188&355,411&$\sim$1185\\
\hline
J.S. Bach Chorales& 382&-&56,441& $\sim$148 \\
\hline
Maestro v3.0.0& 1276 &-&7,040,164&$\sim$5517\\
\hline
\end{tabular}}
\end{table}

\textbf{Evaluation metrics:} 
Several quantitative music metrics were used to evaluate the generated music ~\cite{Metrics,Transformer-GANs}:
PPL (Perplexity, measures the performance of predict the next note);
PCU (Unique pitch classes);
ISR (Nonzero entries in C major scale / Total nonzero entries);
PRS (Time steps where the no. of pitches $\geq$ 4 / Total time steps); 
TUP (Different pitches within a sample); 
PR (Avg. difference of the highest and lowest pitch in semitones); 
APS (Avg. semitone interval between two consecutive pitches);
IOI (Time between two consecutive notes);
PCH (Pitch-Class Histogram Entropy, measures the instability of pitch usage in shorter timescales);
GS (Grooving Pattern Similarity, measures consistency of rhythm across the entire piece);
CPI (Chord Progression Irregularity, measures consistency of harmony across the entire piece); 
SI (Structural Indicator, detects presence of repeated structures within a specified range of timescale);

\textbf{Platform:} 
All models were trained/tested on two Nvidia GeForce RTX 2080-Ti 12 GB GPU.

\subsection{Symbolic Music Generation}
We compared the performance of the present hierarchical model with other outstanding music generation models that had achieved SOTA performance at the moment.
The experiments focused on assessing the pitch diversity (PCU, TUP, PR, APS), non-blank rate (ISR, PRS), rhythm consistency (IOI, PCH, GS), and musical rules (CPI, SI) of the generated samples.
All models were retrained with the same training sets.
We ran the results 5 times and reported the average values.
The minimum unit of a visual token is a semiquaver note.

As shown in Tab. \ref{table2}, our hierarchical model surpasses other models in most metrics.
MusicTransformer and Transformer-XL use the similar relative attention mechanism, and achieve similar results.
Transformer-GANs achieved the better performance in some metrics, especially in short-term music generation (J.S. Bach dataset), duo to adversarial training that reduced the distributional discrepancy between real and generated data.
Our model performs the best result (the value of metrics are the closest to the training set), which can be attributed to the MSN layer that learns the musical style of the training set.
Meanwhile, our model achieved a higher ISR, TUP, PR, and PCH compared to other models, suggesting better pitch diversity and rhythm consistency in our generated samples.
Most excitingly, we achieved the best SI, indicating the superiority of our model in reusing structural melody.

\renewcommand\arraystretch{1.2}
\begin{table*}[!htbp]
\centering
\caption{Ablation studies of FSL and MSN layers on J.S. Bach and Maestro v3.0.0 datasets.
For each metric, the closer to the training set the values is, the better.
The values of the metrics are expected to be close to that for the training set.
The best performance is highlighted in bold.
}
\label{table4}
\resizebox{2\columnwidth}{!}{
\begin{tabular}{|c|c|c|c|c|c|c|c|c|c|c|c|c|c|}
\hline
\textbf{Baseline}&\textbf{Configuration}&\textbf{PPL}&\textbf{PCU}&\textbf{ISR}&\textbf{PRS}&\textbf{TUP}&\textbf{PR}&\textbf{APS}&\textbf{IOI}&\textbf{PCH}&\textbf{GS}&\textbf{CPI}&\textbf{SI}\\
\hline
Training Set (J.S. Bach)&-&-&8.407&0.835&0.422&25.453&34.322&11.671&0.064&2.994&0.930&0.985&0.011\\
\hline
Local Attention&$\left[Global, MSN\right]$&1.872&6.979&0.815&0.417&24.812& 34.832&11.153&0.125&2.92&0.871&0.99&0.009\\
Sparse attention&$\left[Global, MSN\right]$&1.724&\textbf{7.123}&0.817&\textbf{0.419}&24.859&34.699&11.237&0.124&2.934&\textbf{0.922}&\textbf{0.993}&0.013\\ 
\hline
Multi-scale Attention& $\left[L2R, None\right]$&1.917&5.565&0.802&0.391&22.178&41.955&6.983&0.084&2.566&0.879&1&0.005\\
Multi-scale Attention&$\left[L2R, MSN\right]$&1.729&6.933&0.815&0.415&24.68&34.704&\textbf{11.32}&0.127&2.915&0.874&0.997&\textbf{0.011}\\
\hline
Multi-scale Attention&$\left[Global, None\right]$&1.733&6.449&0.789&0.402&20.021&38.186&6.883&\textbf{0.074}&2.501&0.889&0.998&0.003\\
Multi-scale Attention&$\left[Global, MSN\right]$&\textbf{1.716}&7.061&\textbf{0.819}&0.412&\textbf{25.194}&\textbf{34.597}&10.982&0.124&\textbf{2.944}&0.872&0.994&0.01\\ 
\hline
\hline
Training Set (Maestro)&-&-&6.665&0.815&0.429&66.002&68.146&11.598&0.083& 2.886&0.901&0.958&0.208\\
\hline
Local Attention& $\left[Global, MSN\right]$&1.184&6.361&0.786&0.411&52.924&61.422&10.651&0.090&2.817&0.942&0.743&0.089\\
Sparse Attention & $\left[Global, MSN\right]$&1.086&\textbf{6.626}&0.834&0.436&73.965&76.011&10.767&0.082&\textbf{2.863}&0.892&0.998&0.171\\
\hline
Multi-scale Attention &$\left[L2R, None\right]$&1.108&6.516&0.825&0.44&76.0&77.7&\textbf{11.4}&0.087&2.949&0.882&0.997&0.093\\
Multi-scale Attention &$\left[L2R, MSN\right]$&\textbf{1.042}&6.759&0.832&0.435&74.17&76.93&11.09&\textbf{0.083}&2.987&0.887&0.992&0.156\\
\hline
Multi-scale Attention & $\left[Global, None\right]$&1.181&6.751&0.824&0.439&\textbf{66.91}&73.0&11.19&0.084&2.988&0.892&0.994&0.139\\
Multi-scale Attention& $\left[Global, MSN\right]$&1.063&6.716&\textbf{0.815}&\textbf{0.429}&57.26&\textbf{67.62}&11.35&0.087&2.985&\textbf{0.894}&\textbf{0.987}&\textbf{0.191}\\
\hline
\end{tabular}}
\end{table*}

\begin{figure}[htbp]
\centering
\includegraphics[width=0.98\columnwidth]{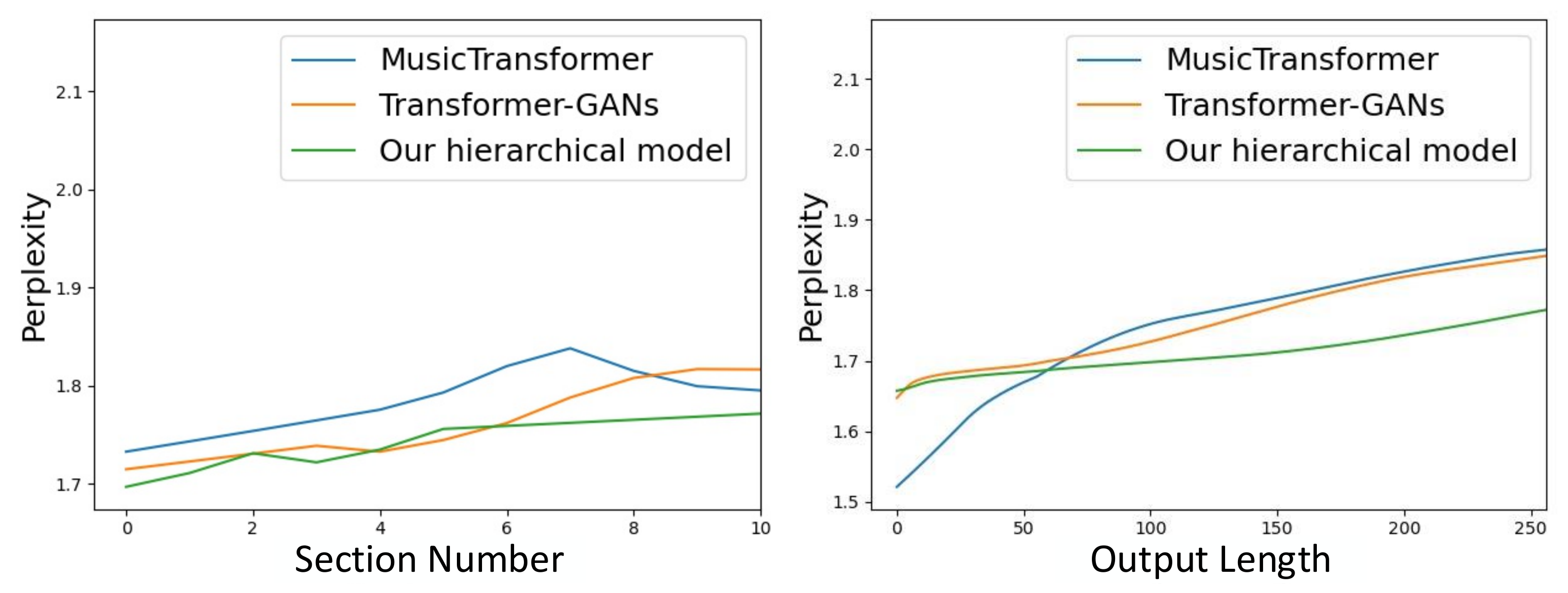}
\centerline{\footnotesize(a) Impact of section number and output length in J.S. Bach dataset}
\includegraphics[width=0.98\columnwidth]{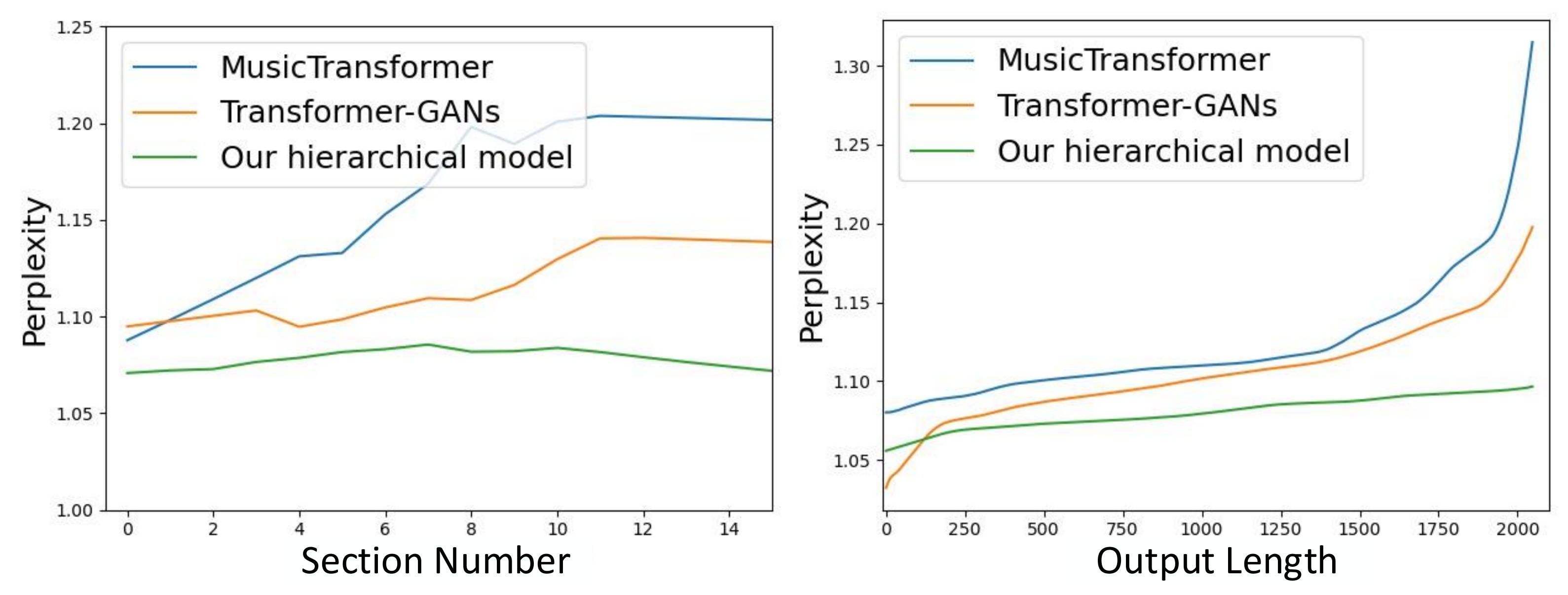}
\centerline{\footnotesize(b) Impact of section number and output length in Maestro dataset}
\caption{Results of the error accumulation analysis.
Note that the PPL of MusicTransformer and Transformer-GANs increases considerably with the length and number of sections, even on the exponential scale.
On the contrary, our model was affected very slightly.
}
\label{fig6}
\end{figure}

\textbf{Error accumulation Analysis:} 
We further tested the generation performance under different conditions, i.e., number of sections and length of the output, using the PPL metric, which measures the underlying performance of the language model.
These tests were performed on the J.S. Bach and Maestro datasets.
As shown in Fig. \ref{fig6}, the PPL increases rapidly when the section number exceeds 6 for MusicTransformer and Transformer-GANs, while staying stable for our hierarchical model. 
Similarly, the PPL for MusicTransformer and Transformer-GANs increases significantly when the output length exceeds 1500, while the drop is much less for our hierarchical model.
This indicates that our model can better avoid the error accumulation problem compared to other Transformer-based models.

\subsection{Ablation studies}
The improvements in the different configurations of the FSL and MSN layers were analyzed by ablation studies.
We performed all the ablation studies on the J.S. Bach and Maestro datasets, which were designed to test three aspects of performance:
(1) The influence of the baseline model choice;
(2) The optimal alignment strategy of candidate windows in the FSL layer; 
(3) The effectiveness of the MSN layer; 
(4) The best configuration of our model.
The results are demonstrated in Tab. \ref{table4}.

The results of ablation studies have suggested the effectiveness of the FSL and MSN layers, while the sparse attention configured with $\left[Global, MSN\right]$ parameters achieved the best performance on most metrics.
Almost all models that use the global strategy (compared to the L2R strategy) to align the candidate windows lead to better performance of music generation. 
The use of the MSN layer resulted in better pitch diversity in the generated music (TUP, PR, and APS are great improvements), validating its utility in musical style control.
These results show the importance of structural segmentation (FSL layer) and music style control (MSN layer) for SMG tasks.

\subsection{Visual Evaluation}
\begin{figure*}[htbp]
\centering
\includegraphics[width=1.8\columnwidth]{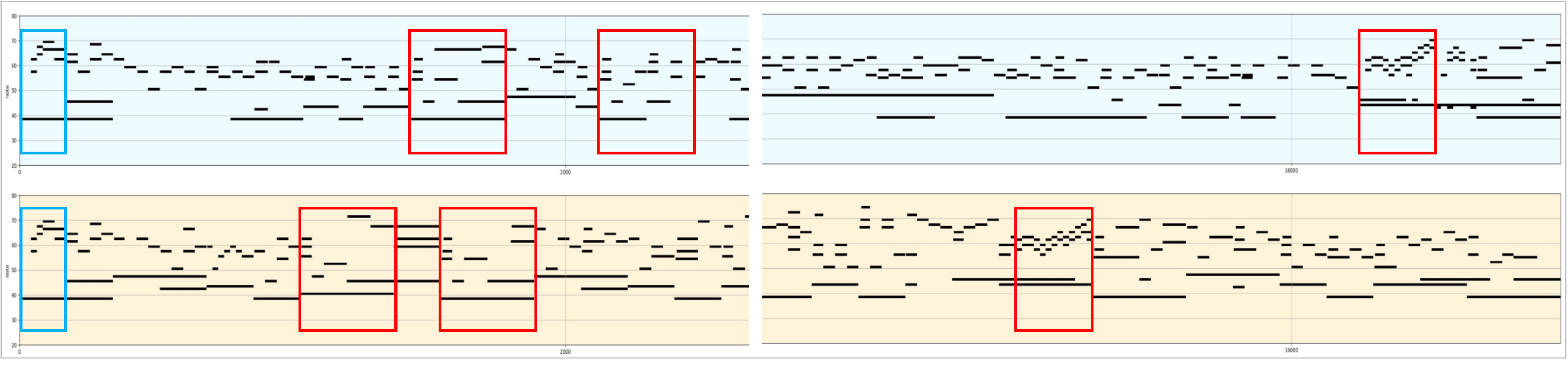}
\caption{Illustration of visual comparison of the generated (upper in shallow blue) and original music (lower in orange). Blue boxes indicate activation notes; Red boxes are examples of reused melody.}
\label{fig7}
\end{figure*}

In order to test the practical performance of our model, we specifically analyzed the structure of the music in the generated samples.
The sample evaluation was demonstrated by the visual MIDI diagram.
Fig. \ref{fig7} shows an example of a composition generated by our model, in which a steady rhythm is maintained compared to the original music, and some strong melodies are recreated (in the red boxes).

\begin{figure*}[htbp]
\centering
\includegraphics[width=1.8\columnwidth]{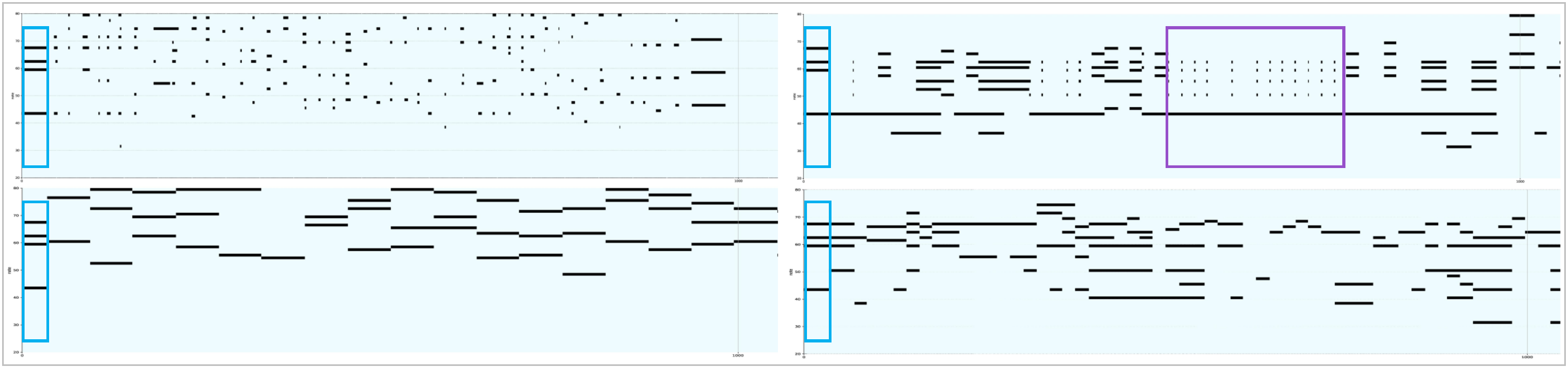}
\caption{Visual comparison of musical samples generated by four different models, i.e., Transformer-XL (top left), MusicTransformer (top right), Transformer-GANs (bottom left), and our hierarchical model (bottom right).
The blue box shows the same activation notes.}
\label{fig8}
\end{figure*}

Further, samples generated by our model and other three comparison models were also presented under the same experimental condition, as shown in Fig. \ref{fig8}. 
In the sample generated by Transformer-XL, all notes are the minimum unit (semiquaver note), making the composition sound monotonous and boring.
The sample generated by MusicTransformer is rhythmically inconsistent because there is an in-equivalent note density in the sequence (see purple box), making the generated music unrealistic.
All notes in the samples generated by Transformer-GANS are large in timescale (longer duration), in higher octaves (higher pitches), and rigid in musical style.
Our model generates samples with lively tonal trends and stable rhythms.
Furthermore, the samples reflect many compositional techniques, such as the use of grace notes (bass tones) and melodic repetition, which conform to the musical sense and are similar to realistic music.

\section{Discussion and Conclusion}
We propose a hierarchical Transformer model to generate music with structural sections and reused chord progressions, which the advantage is learning contextual representation at multi-scale.
Results demonstrated that the present model achieved SOTA performance on two opening datasets.
Ablation studies showed the effectiveness of the proposed FSL and MSN layers.

Another advantage of our model is that the decoding time of our model was reduced to approximately 0.8 times compared to other Transformer models, after using the hierarchical architecture.
For example, the computational complexity of the Sparse Transformer (the most common model) with a maximum length $L$ is $O\left(L \log L\right)$.
In our multi-scale attention, since chord-level and section-level heads avoid connecting all sequence nodes.
Benefiting from parallel decoding, our hierarchical architecture reduces the computational complexity by approximating $O\left(\left(L/m\right) \log \left(L/m\right)\right)$.
Tab. \ref{table5} shows the computational complexity of contemporary canonical models.

\renewcommand\arraystretch{1.2}
\begin{table}[htbp]
\scriptsize 
\centering
\caption{L-related computation complexity. 
$B$ is the number of memory blocks in MusicTransformer.
$m$ is the number of section in the sample. 
}
\label{table5}
\resizebox{0.98\columnwidth}{!}{
\begin{tabular}{|c|c|c|}
\hline
\textbf{Model}&\textbf{Computational Complexity}&\textbf{Decoding Step}\\
\hline
LSTM & $O\left(L\right)$& $L$\\
\hline
Transformer & $O\left(L^{2}\right)$ & $L$\\
\hline
Sparse Transformer & $O\left(L \log L\right)$& $L$\\
\hline
MusicTransformer & $O\left(\left(L/B\right)^{2}\right)$& $L$\\
\hline
Transformer-GANs & $O\left(2L \log L\right)$& $L$\\
\hline
Hierarchical Transformer & $O\left(\left(L/m\right) \log \left(L/m\right)\right)$& $L/m$\\ 
\hline
\end{tabular}}
\end{table}

However, we are also aware of some shortcomings that we have not addressed.
In some cases, the segmentation operation of the FSL layer may lose its effectiveness for dealing with a composition with rhythmic variations, such as in Beethoven's Piano Sonata No.30, 
This problem could be improved by resizing the windows according to the note density of the candidate scope.

In the future, we plan to explore other applications of the hierarchical models, for instance, generating polyphony music that involves multiple tracks.
We can also try other strategies, such as adversarial learning and curriculum learning, to improve the performance of music generation.


\bibliography{anthology}

\begin{thebibliography}{10}
\providecommand{\url}[1]{#1}
\csname url@samestyle\endcsname
\providecommand{\newblock}{\relax}
\providecommand{\bibinfo}[2]{#2}
\providecommand{\BIBentrySTDinterwordspacing}{\spaceskip=0pt\relax}
\providecommand{\BIBentryALTinterwordstretchfactor}{4}
\providecommand{\BIBentryALTinterwordspacing}{\spaceskip=\fontdimen2\font plus
\BIBentryALTinterwordstretchfactor\fontdimen3\font minus
  \fontdimen4\font\relax}
\providecommand{\BIBforeignlanguage}[2]{{%
\expandafter\ifx\csname l@#1\endcsname\relax
\typeout{** WARNING: IEEEtran.bst: No hyphenation pattern has been}%
\typeout{** loaded for the language `#1'. Using the pattern for}%
\typeout{** the default language instead.}%
\else
\language=\csname l@#1\endcsname
\fi
#2}}
\providecommand{\BIBdecl}{\relax}
\BIBdecl

\bibitem{Briot-et-al:1}
\BIBentryALTinterwordspacing
J.~Briot, G.~Hadjeres, and F.~Pachet, \emph{Deep Learning Techniques for Music
  Generation}.\hskip 1em plus 0.5em minus 0.4em\relax Springer, 2020. [Online].
  Available: \url{https://doi.org/10.1007/978-3-319-70163-9}
\BIBentrySTDinterwordspacing

\bibitem{PerformanceRNN}
I.~Simon and S.~Oore, ``Performance rnn: Generating music with expressive
  timing and dynamics,'' \url{https://magenta.tensorflow.org/performance-rnn},
  2017.

\bibitem{VaswaniSPUJGKP17}
A.~Vaswani, N.~Shazeer, N.~Parmar, J.~Uszkoreit, L.~Jones, A.~N. Gomez,
  L.~Kaiser, and I.~Polosukhin, ``Attention is all you need,'' in
  \emph{Advances in Neural Information Processing Systems 30: Annual Conference
  on Neural Information Processing Systems 2017, December 4-9, 2017, Long
  Beach, CA, {USA}}, I.~Guyon, U.~von Luxburg, S.~Bengio, H.~M. Wallach,
  R.~Fergus, S.~V.~N. Vishwanathan, and R.~Garnett, Eds., 2017, pp. 5998--6008.

\bibitem{MusicTransformer}
\BIBentryALTinterwordspacing
C.~A. Huang, A.~Vaswani, J.~Uszkoreit, I.~Simon, C.~Hawthorne, N.~Shazeer,
  A.~M. Dai, M.~D. Hoffman, M.~Dinculescu, and D.~Eck, ``Music transformer:
  Generating music with long-term structure,'' in \emph{7th International
  Conference on Learning Representations, {ICLR} 2019}.\hskip 1em plus 0.5em
  minus 0.4em\relax OpenReview.net, 2019. [Online]. Available:
  \url{https://openreview.net/forum?id=rJe4ShAcF7}
\BIBentrySTDinterwordspacing

\bibitem{Musenet}
C.~M. Payne, ``Musenet,'' \url{https://openai.com/blog/musenet}, 2019.

\bibitem{GPT-2}
A.~Radford, J.~Wu, J.~Clark, A.~Askell, D.~Lansky, D.~Hernandez, and D.~Luan,
  ``Better language models and their implications,''
  \url{https://openai.com/blog/better-language-models/}, 2019, accessed: 2019.

\bibitem{LVM}
\BIBentryALTinterwordspacing
A.~Roberts, J.~H. Engel, C.~Raffel, C.~Hawthorne, and D.~Eck, ``A hierarchical
  latent vector model for learning long-term structure in music,'' in
  \emph{Proceedings of the 35th International Conference on Machine Learning,
  {ICML} 2018, Stockholmsm{\"{a}}ssan, Stockholm, Sweden, July 10-15, 2018},
  ser. Proceedings of Machine Learning Research, J.~G. Dy and A.~Krause, Eds.,
  vol.~80.\hskip 1em plus 0.5em minus 0.4em\relax {PMLR}, 2018, pp. 4361--4370.
  [Online]. Available: \url{http://proceedings.mlr.press/v80/roberts18a.html}
\BIBentrySTDinterwordspacing

\bibitem{MuseGAN}
\BIBentryALTinterwordspacing
H.~Dong, W.~Hsiao, L.~Yang, and Y.~Yang, ``Musegan: Multi-track sequential
  generative adversarial networks for symbolic music generation and
  accompaniment,'' in \emph{Proceedings of the Thirty-Second {AAAI} Conference
  on Artificial Intelligence, (AAAI-18)}.\hskip 1em plus 0.5em minus
  0.4em\relax {AAAI} Press, 2018, pp. 34--41. [Online]. Available:
  \url{https://www.aaai.org/ocs/index.php/AAAI/AAAI18/paper/view/17286}
\BIBentrySTDinterwordspacing

\bibitem{GabrielBSBFC21}
\BIBentryALTinterwordspacing
S.~Gabriel, C.~Bhagavatula, V.~Shwartz, R.~L. Bras, M.~Forbes, and Y.~Choi,
  ``Paragraph-level commonsense transformers with recurrent memory,'' in
  \emph{Thirty-Fifth {AAAI} Conference on Artificial Intelligence, {AAAI}
  2021}.\hskip 1em plus 0.5em minus 0.4em\relax {AAAI} Press, 2021, pp.
  12\,857--12\,865. [Online]. Available:
  \url{https://ojs.aaai.org/index.php/AAAI/article/view/17521}
\BIBentrySTDinterwordspacing

\bibitem{li2019enhancing}
\BIBentryALTinterwordspacing
S.~Li, X.~Jin, Y.~Xuan, X.~Zhou, W.~Chen, Y.~Wang, and X.~Yan, ``Enhancing the
  locality and breaking the memory bottleneck of transformer on time series
  forecasting,'' in \emph{Advances in Neural Information Processing Systems 32:
  Annual Conference on Neural Information Processing Systems 2019, NeurIPS
  2019, December 8-14, 2019, Vancouver, BC, Canada}, H.~M. Wallach,
  H.~Larochelle, A.~Beygelzimer, F.~d'Alch{\'{e}}{-}Buc, E.~B. Fox, and
  R.~Garnett, Eds., 2019, pp. 5244--5254. [Online]. Available:
  \url{https://proceedings.neurips.cc/paper/2019/hash/6775a0635c302542da2c32aa19d86be0-Abstract.html}
\BIBentrySTDinterwordspacing

\bibitem{Reformer}
\BIBentryALTinterwordspacing
N.~Kitaev, L.~Kaiser, and A.~Levskaya, ``Reformer: The efficient transformer,''
  in \emph{8th International Conference on Learning Representations, {ICLR}
  2020}.\hskip 1em plus 0.5em minus 0.4em\relax OpenReview.net, 2020. [Online].
  Available: \url{https://openreview.net/forum?id=rkgNKkHtvB}
\BIBentrySTDinterwordspacing

\bibitem{Transformer-XL}
\BIBentryALTinterwordspacing
Z.~Dai, Z.~Yang, Y.~Yang, J.~G. Carbonell, Q.~V. Le, and R.~Salakhutdinov,
  ``Transformer-xl: Attentive language models beyond a fixed-length context,''
  in \emph{Proceedings of the 57th Conference of the Association for
  Computational Linguistics, {ACL} 2019, Florence, Italy, July 28- August 2,
  2019, Volume 1: Long Papers}, A.~Korhonen, D.~R. Traum, and L.~M{\`{a}}rquez,
  Eds.\hskip 1em plus 0.5em minus 0.4em\relax Association for Computational
  Linguistics, 2019, pp. 2978--2988. [Online]. Available:
  \url{https://doi.org/10.18653/v1/p19-1285}
\BIBentrySTDinterwordspacing

\bibitem{Transformer-GANs}
\BIBentryALTinterwordspacing
A.~Muhamed, L.~Li, X.~Shi, S.~Yaddanapudi, W.~Chi, D.~Jackson, R.~Suresh, Z.~C.
  Lipton, and A.~J. Smola, ``Symbolic music generation with transformer-gans,''
  in \emph{Thirty-Fifth {AAAI} Conference on Artificial Intelligence, {AAAI}
  2021}.\hskip 1em plus 0.5em minus 0.4em\relax {AAAI} Press, 2021, pp.
  408--417. [Online]. Available:
  \url{https://ojs.aaai.org/index.php/AAAI/article/view/16117}
\BIBentrySTDinterwordspacing

\bibitem{WangSW19}
\BIBentryALTinterwordspacing
Q.~Wang, F.~Su, and Y.~Wang, ``A hierarchical attentive deep neural network
  model for semantic music annotation integrating multiple music
  representations,'' in \emph{Proceedings of the 2019 on International
  Conference on Multimedia Retrieval, {ICMR} 2019}.\hskip 1em plus 0.5em minus
  0.4em\relax {ACM}, 2019, pp. 150--158. [Online]. Available:
  \url{https://doi.org/10.1145/3323873.3325031}
\BIBentrySTDinterwordspacing

\bibitem{Segmentation}
\BIBentryALTinterwordspacing
M.~C. McCallum, ``Unsupervised learning of deep features for music
  segmentation,'' in \emph{{IEEE} International Conference on Acoustics, Speech
  and Signal Processing, {ICASSP} 2019}.\hskip 1em plus 0.5em minus 0.4em\relax
  {IEEE}, 2019, pp. 346--350. [Online]. Available:
  \url{https://doi.org/10.1109/ICASSP.2019.8683407}
\BIBentrySTDinterwordspacing

\bibitem{Hierarchical:2}
\BIBentryALTinterwordspacing
S.~Dai, Z.~Jin, C.~Gomes, and R.~B. Dannenberg, ``Controllable deep melody
  generation via hierarchical music structure representation,'' in
  \emph{Proceedings of the 22nd International Society for Music Information
  Retrieval Conference, {ISMIR} 2021, Online, November 7-12, 2021}, J.~H. Lee,
  A.~Lerch, Z.~Duan, J.~Nam, P.~Rao, P.~van Kranenburg, and A.~Srinivasamurthy,
  Eds., 2021, pp. 143--150. [Online]. Available:
  \url{https://archives.ismir.net/ismir2021/paper/000017.pdf}
\BIBentrySTDinterwordspacing

\bibitem{HuangB17}
\BIBentryALTinterwordspacing
X.~Huang and S.~J. Belongie, ``Arbitrary style transfer in real-time with
  adaptive instance normalization,'' in \emph{{IEEE} International Conference
  on Computer Vision, {ICCV} 2017, Venice, Italy, October 22-29, 2017}.\hskip
  1em plus 0.5em minus 0.4em\relax {IEEE} Computer Society, 2017, pp.
  1510--1519. [Online]. Available: \url{https://doi.org/10.1109/ICCV.2017.167}
\BIBentrySTDinterwordspacing

\bibitem{RA_AIN}
\BIBentryALTinterwordspacing
J.~Ling, H.~Xue, L.~Song, R.~Xie, and X.~Gu, ``Region-aware adaptive instance
  normalization for image harmonization,'' in \emph{{IEEE} Conference on
  Computer Vision and Pattern Recognition, {CVPR} 2021, virtual, June 19-25,
  2021}.\hskip 1em plus 0.5em minus 0.4em\relax Computer Vision Foundation /
  {IEEE}, 2021, pp. 9361--9370. [Online]. Available:
  \url{https://openaccess.thecvf.com/content/CVPR2021/html/Ling\_Region-Aware\_Adaptive\_Instance\_Normalization\_for\_Image\_Harmonization\_CVPR\_2021\_paper.html}
\BIBentrySTDinterwordspacing

\bibitem{ChawlaY20}
\BIBentryALTinterwordspacing
K.~Chawla and D.~Yang, ``Semi-supervised formality style transfer using
  language model discriminator and mutual information maximization,'' in
  \emph{Findings of the Association for Computational Linguistics: {EMNLP}
  2020, Online Event, 16-20 November 2020}, ser. Findings of {ACL}, T.~Cohn,
  Y.~He, and Y.~Liu, Eds., vol. {EMNLP} 2020.\hskip 1em plus 0.5em minus
  0.4em\relax Association for Computational Linguistics, 2020, pp. 2340--2354.
  [Online]. Available:
  \url{https://doi.org/10.18653/v1/2020.findings-emnlp.212}
\BIBentrySTDinterwordspacing

\bibitem{KusnerH16}
\BIBentryALTinterwordspacing
M.~J. Kusner and J.~M. Hern{\'{a}}ndez{-}Lobato, ``{GANS} for sequences of
  discrete elements with the gumbel-softmax distribution,'' \emph{CoRR}, vol.
  abs/1611.04051, 2016. [Online]. Available:
  \url{http://arxiv.org/abs/1611.04051}
\BIBentrySTDinterwordspacing

\bibitem{MIboundary}
B.~Poole, S.~Ozair, A.~van~den Oord, A.~Alemi, and G.~Tucker, ``On variational
  bounds of mutual information,'' in \emph{Proceedings of the 36th
  International Conference on Machine Learning, {ICML}}, 2019, pp. 5171--5180.

\bibitem{DeepBach}
\BIBentryALTinterwordspacing
G.~Hadjeres, F.~Pachet, and F.~Nielsen, ``Deepbach: a steerable model for bach
  chorales generation,'' in \emph{Proceedings of the 34th International
  Conference on Machine Learning, {ICML} 2017, Sydney, NSW, Australia, 6-11
  August 2017}, ser. Proceedings of Machine Learning Research, D.~Precup and
  Y.~W. Teh, Eds., vol.~70.\hskip 1em plus 0.5em minus 0.4em\relax {PMLR},
  2017, pp. 1362--1371. [Online]. Available:
  \url{http://proceedings.mlr.press/v70/hadjeres17a.html}
\BIBentrySTDinterwordspacing

\bibitem{MAESTRO}
\BIBentryALTinterwordspacing
C.~Hawthorne, A.~Stasyuk, A.~Roberts, I.~Simon, C.~A. Huang, S.~Dieleman,
  E.~Elsen, J.~H. Engel, and D.~Eck, ``Enabling factorized piano music modeling
  and generation with the {MAESTRO} dataset,'' in \emph{7th International
  Conference on Learning Representations, {ICLR} 2019, New Orleans, LA, USA,
  May 6-9, 2019}.\hskip 1em plus 0.5em minus 0.4em\relax OpenReview.net, 2019.
  [Online]. Available: \url{https://openreview.net/forum?id=r1lYRjC9F7}
\BIBentrySTDinterwordspacing

\bibitem{Metrics}
\BIBentryALTinterwordspacing
S.~Wu and Y.~Yang, ``The jazz transformer on the front line: Exploring the
  shortcomings of ai-composed music through quantitative measures,'' in
  \emph{Proceedings of the 21th International Society for Music Information
  Retrieval Conference, {ISMIR} 2020, Montreal, Canada, October 11-16, 2020},
  J.~Cumming, J.~H. Lee, B.~McFee, M.~Schedl, J.~Devaney, C.~McKay,
  E.~Zangerle, and T.~de~Reuse, Eds., 2020, pp. 142--149. [Online]. Available:
  \url{http://archives.ismir.net/ismir2020/paper/000339.pdf}
\BIBentrySTDinterwordspacing

\end{thebibliography}
\bibliographystyle{IEEEtran}

\begin{figure*}[htbp]
\centering
\includegraphics[width=1.98\columnwidth]{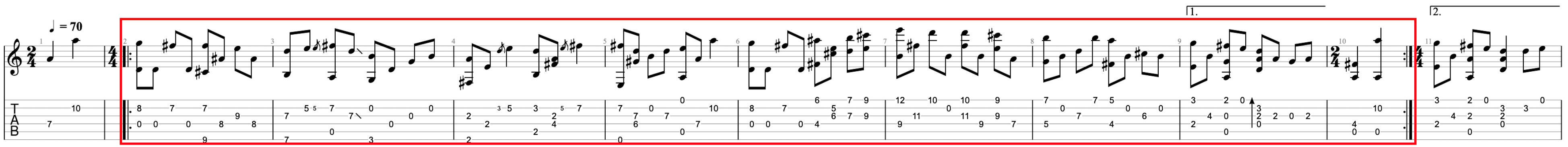}
\centerline{\footnotesize(a) The GTP data}
\includegraphics[width=1.98\columnwidth]{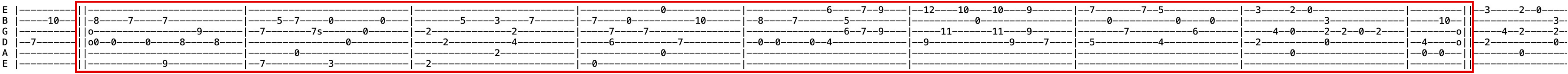}
\centerline{\footnotesize(b) The ASCII data}
\includegraphics[width=1.98\columnwidth]{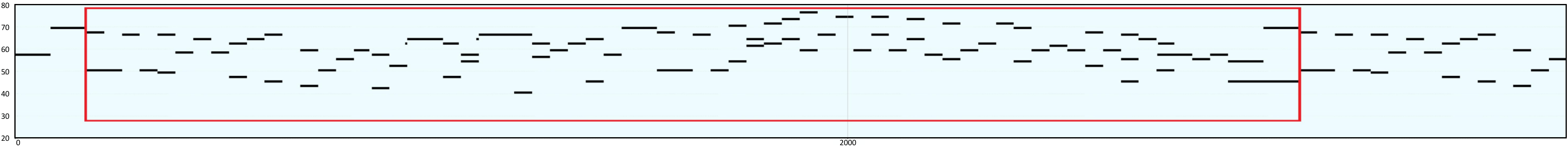}
\centerline{\footnotesize(c) The MIDI data}
\caption{
An example of a music section in GTP data with the corresponding ASCII data and MIDI data.
The section label was extracted in the GTP data and noted on the MIDI data through the ASCII code transfer method.}
\label{fig9}
\end{figure*}

\appendix
\textbf{Data Preprocessing:} 
To pre-train the FSL layer in a supervised manner, the location and length of structural section are needed, which can be independently extracted from the GuitarPro (GTP) software. 
We proposed a transfer method for automatically marking the section regions from GTP data. 
The approach is as follows: 
(1) The GTP data was first converted into MIDI data and the corresponding ASCII text was exported;
(2) The ASCII symbols were discretized onto a note grid and then serialized by iterating through all the symbols within a time step;
(3) The section labels can be marked by adding a location coordinate since there is a direct correspondence between the sequence location and the grid location. 
An example of a music section in GTP data with the corresponding ASCII data and MIDI data is illustrated in Fig. \ref{fig9}.

\begin{figure*}[htbp]
\centering
\includegraphics[width=1.96\columnwidth]{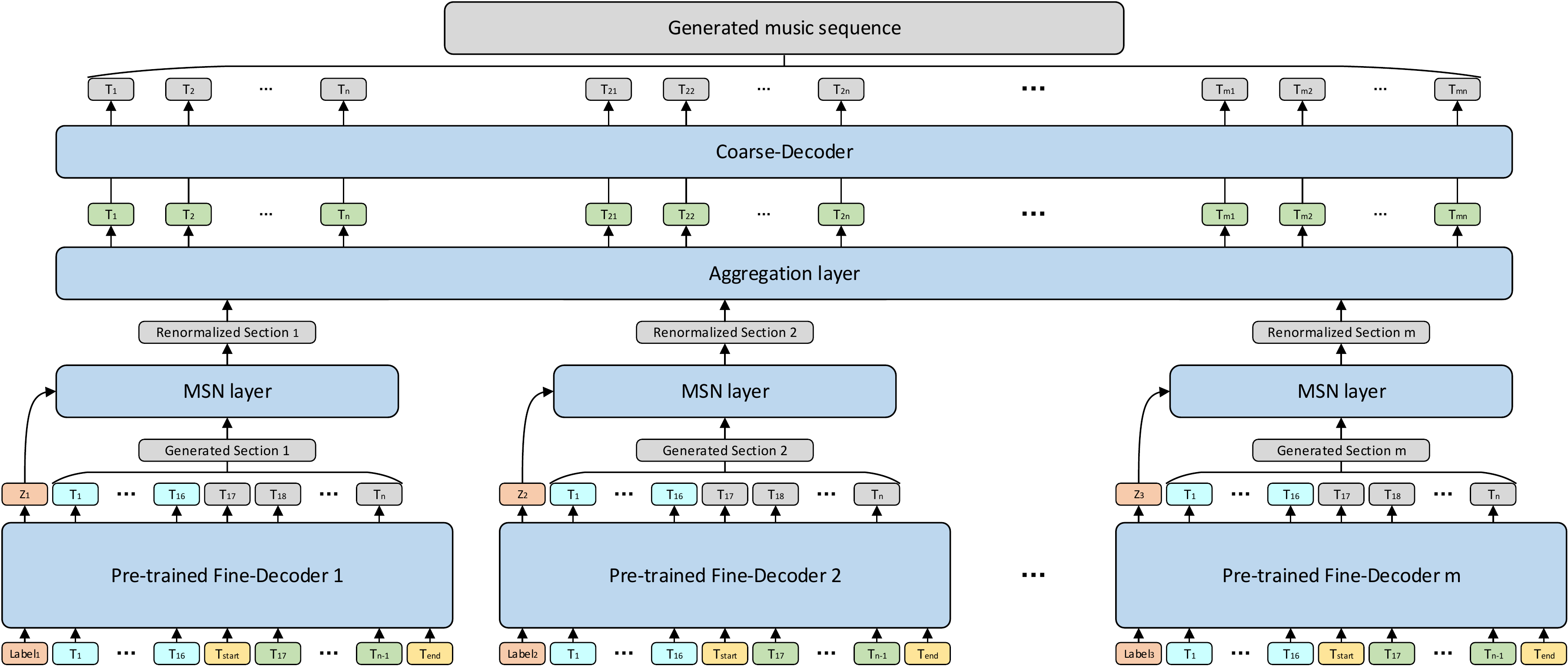}
\caption{Illustration of the training details. 
Indigo boxes represent the starting notes (16 semiquaver notes), which is used to activate the initial state in each section;
Green boxes and Gray boxes represent the input and output of the decoder, respectively; 
Orange boxes represent section labels ($label_{i}$) and style representation ($z$) of the generated section.
Yellow boxes represent start token and end token, which are used to change the state of the model.
}
\label{fig10}
\end{figure*}

In GTP data, the symbols ``$\|:$" and ``$:\|$" represent the beginning and the end of the section, respectively, and are equivalent to the symbol ``$\|$" in ASCII data.
These section labels can be synchronously marked in the MIDI data. 
We have compiled 3188 sections from various genres of music, including blues, flamenco, rock, classical, and other music. 
The diversity of music styles in the music samples enables the FSL layer to generalize better and learn to fragment musics with various styles.
In the FSL layer, the number $K$ of candidate windows was set to 8, and the window size was selected from $[8, 16, 32, 64, 128, 256, 512, 1024]$ to accommodate music sections with various musical styles.

\textbf{Training details:} 
Applying LMs to downstream tasks by pre-training and fine-tuning is a common strategy.
Since the Transformer-based model has the problem of error accumulation when dealing with musical sequences with thousands of notes, it is also slow.
We first pre-train fine-decoders to generate sections at a fine scale and then fine-tune the combined sequence at a coarse scale.
The coarse-encoder and fine-decoders were trained with different default hyper-parameters.
While maintaining the overall design of our hierarchical model, we set the maximum length of the coarse-decoder to be 4096, as in other Transformer-based models in comparison.
The maximum length of fine-decoders is chosen from [256, 512, 768, 1024], using a carry-up strategy to fit the section length (the smallest one that exceeds the section length).

Then, we fix the weights of fine-decoders and train the MSN layer and the coarse decoder.
The the MSN layer is optimized together with the coarse-decoder.
The MI-based loss $L_{style}$ is used to optimize the coarse-decoder, which is calculated as the MI of the input and the renormalized music sections.
The training process of coarse-decoder is shown in Fig. \ref{fig10}.

In our experiments, both fine-decoders and coarse-decoder have 6 hidden layers.
We implemented the model in TensorFlow framework and the hyper-parameters for training were as follows:
\begin{itemize}
\item (a) 1e-03 initial learning rate minimized with 1e-04 weight decay; 
\item (b) 100 epoch and 8 batch size; 
\item (c) 0.2 dropout; 
\item (d) dynamic position embedding;
\item (e) multiple GPU training and early stopping strategy.
\end{itemize}
\end{document}